\documentclass[amssymb,amsmath,prl,twocolumn,superscriptaddress,nofootinbib]{revtex4-1}

\usepackage{graphicx}
\usepackage{dcolumn}
\usepackage{bm}
\usepackage{graphicx}
\usepackage{array}
\usepackage{booktabs}
\usepackage {xcolor}
\usepackage{amssymb}
\usepackage{pifont}

\usepackage{multirow}
\usepackage{diagbox}
\usepackage{comment}
\usepackage{enumerate}
\usepackage{booktabs}
\usepackage[normalem]{ulem}
\usepackage[utf8]{inputenc}
\usepackage{hyperref}
\usepackage{units}
\usepackage{enumerate}
\usepackage{amsmath}
\hypersetup{
    colorlinks = true,
    linkcolor = {blue},
    citecolor = {blue},
    urlcolor = {blue},
    linkbordercolor = {white},
    }
\usepackage{xr}

\newcommand{\newcontent}[1]{{\textcolor{black}{#1}}}

\begin{document}

\def\nrtidalvthree{{\texttt{NRTidalv3}}}
\def\nrtidal{{\texttt{NRTidal}}}
\def\nrtidalvtwo{{\texttt{NRTidalv2}}}
\def\imrphenomxas{{\texttt{IMRPhenomXAS}}}
\def\imrphenomxasnrtidalthree{{\texttt{IMRPhenomXAS\_NRTidalv3}}}
\def\imrphenomxasninePN{{\texttt{IMRPhenomXAS\_9PNTidal}}}
\def\lalsuite{{\texttt{LALSuite}}}
\def\kyototidal{{\texttt{KyotoTidal}}}

\newcommand{\AEI}{\affiliation {Max Planck Institute for Gravitational Physics (Albert Einstein Institute), Am M\"uhlenberg 1, Potsdam 14476, Germany}}
\newcommand{\UP}{\affiliation {Institut f\"{u}r Physik und Astronomie, Universit\"{a}t Potsdam, Haus 28, Karl-Liebknecht-Str. 24/25, 14476, Potsdam, Germany}}

\title{Data-driven approach for extracting tidal information from neutron star binary mergers observed with the Einstein Telescope}

\author{Adrian Abac}
\email{adrian.abac@aei.mpg.de}
\AEI \UP

\author{Anna Puecher}
\UP 

\author{Jonathan Gair}
\AEI

\author{Tim Dietrich}
\UP \AEI

\date{\today}

\begin{abstract}
The recent breakthroughs regarding the detection of compact binary mergers via gravitational waves opened up a new window to the Universe. Gravitational-wave models have been essential to this success since they are necessary to infer the properties of the compact binary system from the observational data. Next-generation detectors, such as the Einstein Telescope, will allow for more observations of binary neutron star mergers with higher precision, making accurate waveform models crucial in describing these systems. In this article, we propose a novel approach for constructing phenomenological waveform models informed by observational data. Using mock data representing a one-year operation of the Einstein Telescope as our baseline, we demonstrate how the results improve as more events are included in the calibration. This method offers a new and complementary approach for developing sophisticated gravitational-wave models compared to classical techniques that employ analytical computations and numerical-relativity simulations. Improved waveform models will then yield more accurate parameter estimation. 
\end{abstract}

\maketitle

\paragraph{Introduction.---}\label{section: introduction}
The discovery of the first binary neutron star (BNS) merger GW170817~\cite{LIGOScientific:2017vwq} by the Advanced LIGO~\cite{LIGOScientific:2014pky} and Advanced VIRGO~\cite{VIRGO:2014yos} detectors opened a new chapter in multi-messenger astronomy. Observations of gravitational-wave (GW) signals from BNSs~\cite{LIGOScientific:2020aai}, potentially in conjunction with their electromagnetic counterparts~\cite{LIGOScientific:2017ync, Rastinejad:2022zbg}, contribute to the understanding of the equation of state (EOS) governing supranuclear-dense matter, e.g.,~\cite{Ozel:2016oaf, Huth:2021bsp, Burgio:2021vgk, Alford:2022bpp, Zhu:2023ijx, Lattimer:2021emm, Koehn:2024set} for recent reviews, to an independent measurement of the expansion rate of our Universe, e.g.,~\cite{LIGOScientific:2017adf}, and allow us to study the formation of heavy elements, e.g., \cite{Rosswog:2017sdn,Watson:2019xjv}.

The analysis of these observations involves matching the GW signals with theoretical predictions using a particular waveform model chosen a priori. These models are typically based on the Post-Newtonian (PN) expansion~\cite{Vines:2011ud, Damour:2012yf, Blanchet:2013haa, Henry:2020ski, Narikawa:2023deu, Mandal:2024iug, Dones:2024odv}, which is a perturbative approach valid for small velocities and large separations; on the effective-one-body (EOB) formalism, which maps the two-body problem into a single test particle in an effective metric~\cite{Buonanno:1998gg, Buonanno:2000ef, Damour:2009zoi, Bohe:2016gbl, Hotokezaka:2015xka, Hinderer:2016eia,Steinhoff:2016rfi,Akcay:2018yyh,Nagar:2018plt,Bernuzzi:2014owa,Dietrich:2017feu,Nagar:2018gnk, Gamba:2023mww, Gamba:2022mgx, Lackey:2016krb, Lackey:2018zvw, Purrer:2014fza, Mihaylov:2021bpf, Tissino:2022thn, Gamba:2020ljo, Nagar:2018gnk}; or on a phenomenological approach~\cite{Kawaguchi:2018gvj,Dietrich:2017aum,Dietrich:2018uni,Dietrich:2019kaq,Abac:2023ujg, Colleoni:2025aoh, Williams:2024twp} that takes PN and/or EOB information and, to cover the final orbits before merger, calibrates their extensions to numerical-relativity (NR) simulations~\cite{Hotokezaka:2015xka, Kawaguchi:2018gvj, Kiuchi:2019kzt, Foucart:2018lhe, Dietrich:2018phi, Ujevic:2022qle, Gonzalez:2022mgo}. 

Unlike binary black hole (BBH) systems, for BNSs, a significant contribution to the phase, particularly in the late inspiral, arises from the presence of finite-size or tidal effects resulting from the deformation of one neutron star as it responds to the external gravitational field of its companion. The measure of this deformation is encoded in the tidal deformability parameter $\Lambda = (2/3)k_2C^{-5}$, where $k_2$ is the gravitational Love number (with typical values around 0.2 to 0.3 depending on the EOS and the star's mass), and $C = M/R$ is the compactness of the star, with $R$ being the radius and $M$ being the mass~\cite{Hinderer:2007mb, Flanagan:2007ix, Vines:2011ud, Chatziioannou:2024tjq}. 

Tidal effects enter the phase at the 5PN order (i.e., proportional to $(v/c)^{10}$) and are known analytically up to 7.5PN~\cite{Damour:2012yf, Henry:2020ski, Narikawa:2023deu, Mandal:2024iug, Dones:2024odv}. In this work, we will construct a waveform with a simple, extended {\it pseudo}-PN tidal part up to 9PN, whose unknown higher-order PN coefficients are left as free parameters and are inferred through Bayesian inference from the observed data, together with the source parameters.
The higher-order coefficients (HOCs) allow a flexible representation of the tidal effects. Going to an even higher PN representation would generally increase the flexibility and accuracy of the model (\newcontent{see the supplementary material~\cite{supplementary_material}}), but would also increase the number of free parameters such that more data would be needed to obtain sensible information. For this reason, we restrict the current work to HOCs up to the 9PN order, which introduces six parameters in addition to the unknown parameters that describe the other properties of the observed sources, e.g., masses, spins, luminosity distance, and inclination. 

Traditionally, waveform model development has involved a combination of improving the analytical knowledge, e.g., through computing higher PN orders~\cite{Hinderer:2007mb,Vines:2011ud,Damour:2012yf,Henry:2020ski, Narikawa:2023deu, Mandal:2024iug, Dones:2024odv}, and advances in NR simulations~\cite{Baiotti:2010xh,Bernuzzi:2012ci,Hotokezaka:2013mm,Dietrich:2017feu}. While NR simulations solve the Einstein's Field Equations directly, their use is limited by their short duration (i.e., they only generate waveforms in the late inspiral), computational cost, finite resolution, and the complexity when simulating generic systems (e.g., precessing or highly unequal mass mergers). For this reason, it is unclear whether advances in NR would ensure that the target accuracy necessary to interpret GW observations made by the third generation of detectors would be reached, e.g.,~\cite{Kunert:2021hgm,Branchesi:2023mws, Gamba:2020wgg}.
For this purpose, we propose a new method for reconstructing the tidal phase using observed data from the Einstein Telescope (ET)~\cite{Maggiore:2019uih,ET_2020,ET_2024, Branchesi:2023mws, Abac:2025saz}. \newcontent{In this work, we demonstrate that we can leverage observational data to extract this tidal information, serving as an alternative avenue to traditional GW modeling approaches using analytical methods and NR calibration. Moreover, it is distinct from real-time data-driven methods in other fields such as robotics and state estimation, e.g.,~\cite{Kalman:1960, Meiss:2005}.}

Throughout this work, we assume geometric units $G = c = 1$. $M_A$ denotes the mass of the primary component with tidal deformability $\Lambda_A$, and $M_B$ is the secondary mass with tidal deformability $\Lambda_B$. Each star has a dimensionless aligned spin component denoted by $\chi_A$ and $\chi_B$, respectively. The mass ratio is defined as $q = M_A/M_B \ge 1$. The total mass is then $M = M_A + M_B$, and the normalized mass is $X_{A,B} = M_{A,B}/M$. The chirp mass is defined as $\mathcal{M}_c = (M_A M_B)^{3/5}/M^{1/5}$. 

\paragraph{Methodology.---}\label{section: Methodology}

For the creation of our mock data, we assume as our default setup a triangular ET ($\Delta$), with 10-km arm-length~\cite{et_psd} and placed in Limburg~\cite{Maggiore:2019uih,ET_2020,ET_2024, Branchesi:2023mws}. We have cross-checked our results using the alternative proposed design for ET consisting of two L-shaped detectors (2L) with 15-km arm-length~\cite{et_psd} (placed in Sardinia and Lusatia), and when employing a 3G network of a triangular ET~\cite{Hild:2010id} and the Cosmic Explorer~\cite{Evans:2021gyd, CE-T2000017-v8}. The results of both analyses are presented in the supplementary material~\cite{supplementary_material}, and some important findings are also referred to in the main text.

We simulate GW events from a catalog of 1000 sources with masses assumed to follow the \texttt{FLAT\_Q} model distribution of Ref.~\cite{Landry:2021hvl} (see supplementary material~\cite{supplementary_material}). The corresponding $\Lambda$ is computed using as a common EOS, the EOS from Ref.~\cite{Koehn:2024set} yielding the largest likelihood when comparing with current observational data and nuclear physics computations. 
From these, we further select the $N = 100$ events with the highest signal-to-noise ratio (SNR).   
Following this procedure, the catalog's SNR ranges from 74 to 456, with 99\% events below SNR = 250, and 58\% below SNR = 100, consistent with the expected number of cumulative BNS events detected per year by $\Delta$-ET~\cite{Branchesi:2023mws}.

The simulated signals are produced with the waveform model \imrphenomxasnrtidalthree, which employs \imrphenomxas~\cite{Pratten:2020fqn} as its BBH baseline and incorporates the \nrtidalvthree~\cite{Abac:2023ujg, lalsuite} tidal description. We employ an effective or pseudo-9PN extension to the 7.5PN tidal phase, whose HOCs will be inferred during parameter estimation. In the frequency domain, the 9PN and \nrtidalvthree\ tidal phases are represented by the general form 
\begin{equation}
    \psi_T(x) = -c_{\rm Newt}^A \kappa_A(x) x^{5/2}P_A(x) + \left[A\leftrightarrow B\right],
\end{equation}
where $c_{\rm Newt}^{A,B}$ is the leading order PN constant, $\kappa_{A,B}(x)$ is the dynamical tidal parameter, and $P_{A,B}(x)$ is a rational function (usually a polynomial for the PN representation, a Padè approximant for \nrtidalvthree) of the PN parameter or frequency $x = (\pi M f)^{2/3}$.
More details about these waveform approximants can be found in the supplementary material~\cite{supplementary_material}. 

Throughout this work, we assume that the BBH baseline is well-modeled and sufficiently accurate. This assumption, which also underlies the investigation of the effect of dynamical tides in Ref.~\cite{Pratten:2021pro}, is justified by the fact that BBH waveforms can already be calibrated with extremely accurate NR simulations~\cite{Boyle:2019kee}. As for phenomenological models, we attach the 9PN phase to the BBH waveform model. Here we use the spin-aligned \imrphenomxas\ model, with the six HOCs (three for each star) as additional free parameters. The approach of extending the phase with the HOCs is reminiscent of the framework of parametrized tests of general relativity (GR)~\cite{Li:2011cg,Agathos:2013upa, Meidam:2017dgf, Mehta:2022pcn, Sanger:2024axs, LIGOScientific:2018dkp, Sanger:2024axs}. However, in this work, we assume that GR is the correct theory of gravity.

For the inference of the tidal phase, we perform Bayesian parameter estimation (PE) using \texttt{bilby}~\cite{Ashton:2018jfp, Smith:2019ucc}, and the \texttt{dynesty}~\cite{Speagle:2019ivv} sampler. Bayes theorem reads $p(\vec{\theta} | \mathbf{d}, {\Omega}) \propto \mathcal{L}(\mathbf{d}|\vec{\theta}, {\Omega})\pi(\vec{\theta}|{\Omega})$, where $p(\vec{\theta} | \mathbf{d}, {\Omega})$ is the posterior probability distribution of source parameters $\vec{\theta}$ given the data $\mathbf{d}$ and model $\Omega$, $\pi(\vec{\theta}|{\Omega})$ is the prior probability distribution, and $\mathcal{L}(\mathbf{d}|\vec{\theta}, {\Omega})$ the likelihood of obtaining $\mathbf{d}$ given parameters $\vec{\theta}$ under the model $\Omega$. To reduce the computational cost of the analysis, the likelihood is calculated with the adaptive frequency resolution method~\cite{Morisaki:2021ngj}, \newcontent{which divides the frequency range of the GW signal into multiple frequency bands; cf.~~\cite{supplementary_material} for the validation of the method in our analysis.}
The distributions used in building the catalog and the prior distributions for each parameter are provided in the supplementary material~\cite{supplementary_material}.
We set the minimum frequency to $f_{\rm min} = 5\, \rm Hz$, and limit the analysis of the tidal phase up to the merger of the two stars, where existing tidal models such as \imrphenomxasnrtidalthree\ are expected to be applicable.

\begin{figure}[t]
\centering
\includegraphics[width=\linewidth]{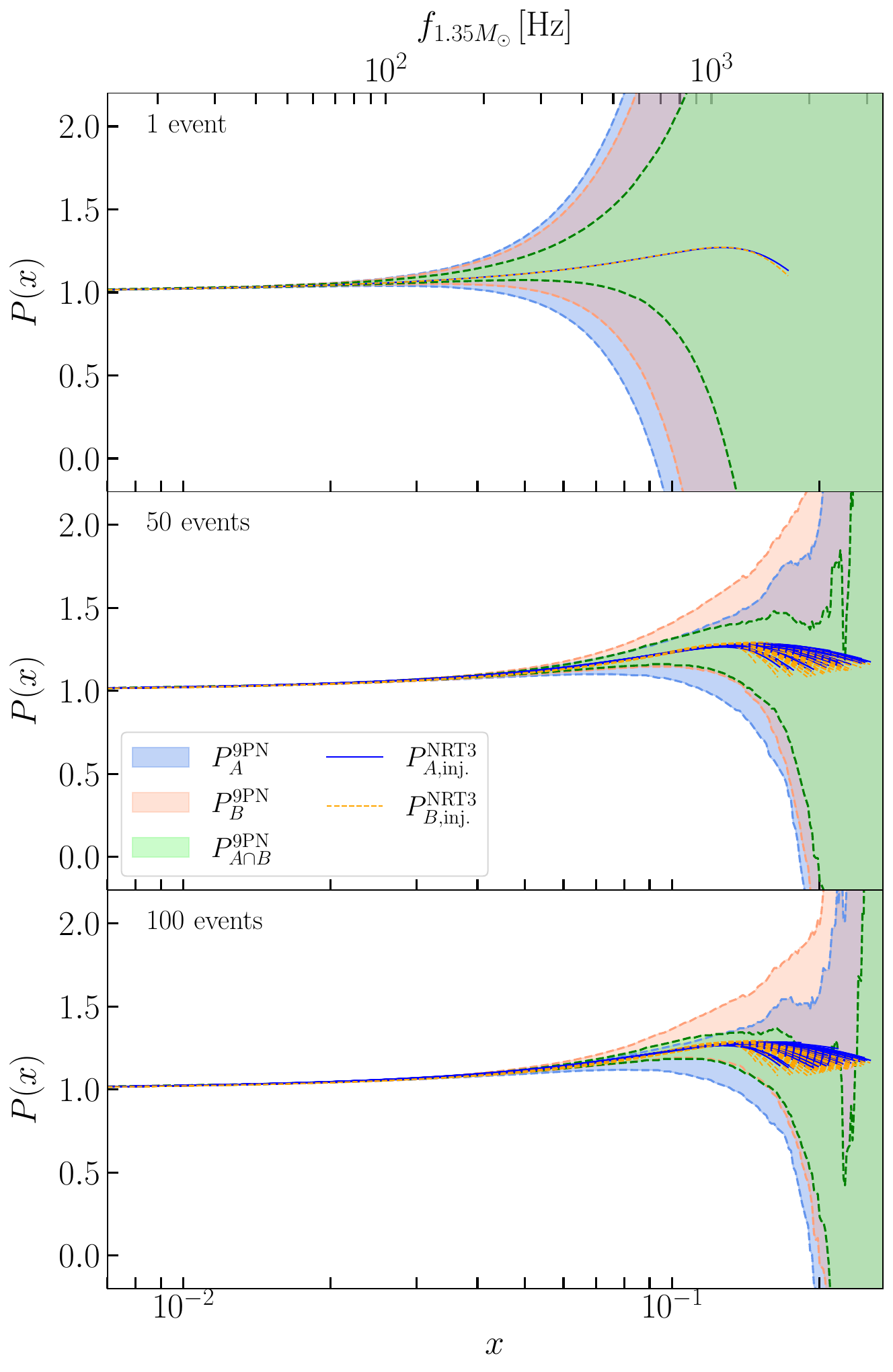}
\caption{The $P(x)$ bands for $N=1$, 50, and 100 events. This includes individual contributions from stars $A$ and $B$ for each event as well as their combined contributions from multiplying their joint posteriors. The curves for \nrtidalvthree\ used for the injection are also plotted, up to the merger frequency. We also include a secondary scale for the frequency $f_{1.35M_\odot}$ for a canonical equal-mass system with component mass $M_{A,B} = 1.35 M_{\odot}$ at the top.}
\label{fig: phaseplot_allevents}
\end{figure}

From the PE analysis, we obtain a posterior distribution over the source parameters and HOCs for each event. Sampling over these posterior distributions, we reconstruct the posterior for the 9PN polynomials $P_{A}^{\rm 9PN}$ and $P_{B}^{\rm 9PN}$, effectively inferring the posterior distribution for $P^{\rm 9PN}(x)$. Similarly, from the parameters' prior, we reconstruct the prior on $P_{A}^{\rm 9PN}$ and $P_{B}^{\rm 9PN}$ for each observation. Assuming that the variation of $P(x)$ from the spread in the modeled population is smaller than the uncertainty of the measurement (cf.~bottom panel of Fig.~\ref{fig: phaseplot_allevents}), we consider a common $P(x)$ when computing joint posteriors from multiple events by multiplying the individual posteriors normalized to their corresponding prior. Since both stars in the binary contribute to the full tidal signal, we can finally compute the joint posteriors for stars A and B to obtain $P_{A\cap B}^{\rm 9PN}$, which effectively doubles the available information. The joint posteriors are obtained by computing the normalized posterior of the average value of $P_A(x)$ and $P_B(x)$. \newcontent{This entire approach of combining posteriors from multiple events effectively constructs a model of $P(x)$ (and, ultimately, of the tidal contributions to the waveform), that is fully informed by the data.} 

\paragraph{Results.---}\label{section: Results}

Figure~\ref{fig: phaseplot_allevents} shows the posterior on the tidal phase function $P(x)$ when combining an increasing number of events, randomly selected among the 100 analyzed systems. We show 90\% confidence intervals for the posterior corresponding to each star, $P_A^{\rm 9PN}$ and $P_B^{\rm 9PN}$, and to the joint one, $P_{A\cap B}^{\rm 9PN}$, as well. For completeness, we also plot the \nrtidalvthree\ Padè functions used to simulate the signals, $P_{A,\rm inj}^{\rm NRT3}$ and $P_{B,\rm inj}^{\rm NRT3}$.
For a single event, the confidence interval is relatively wide, in particular for large $x$, reflecting the uncertainty in the PE results for a single observation. However, we find that the injected values lie inside the inferred confidence intervals. 

For multiple events, the confidence bands for $P(x)$ become narrower, especially at large $x$, further constraining $P(x)$ while still being centered around the injected \nrtidalvthree\ ones. As expected, the tightest constraint for $P(x)$ is achieved when all 100 events are combined. For $x \gtrsim 0.2$ (which, for a canonical equal-mass system with component masses $M_{A,B} = 1.35 M_{\odot}$, corresponds to $f_{1.35M_{\odot}} \gtrsim 2000 \,\rm Hz$), the band for $P_{A\cap B}$ deviates from some of the injected values. This is attributed to the fact that different events terminate at different merger frequencies. In general, our findings imply that more observational data will lead to tighter constraints on $P(x)$ and, therefore, to the extraction of the tidal information it represents. Moreover, the consistency of this result also proves the robustness of this approach when data from multiple sources are combined. 

\begin{figure}[t]
\centering  \includegraphics[width=\linewidth]{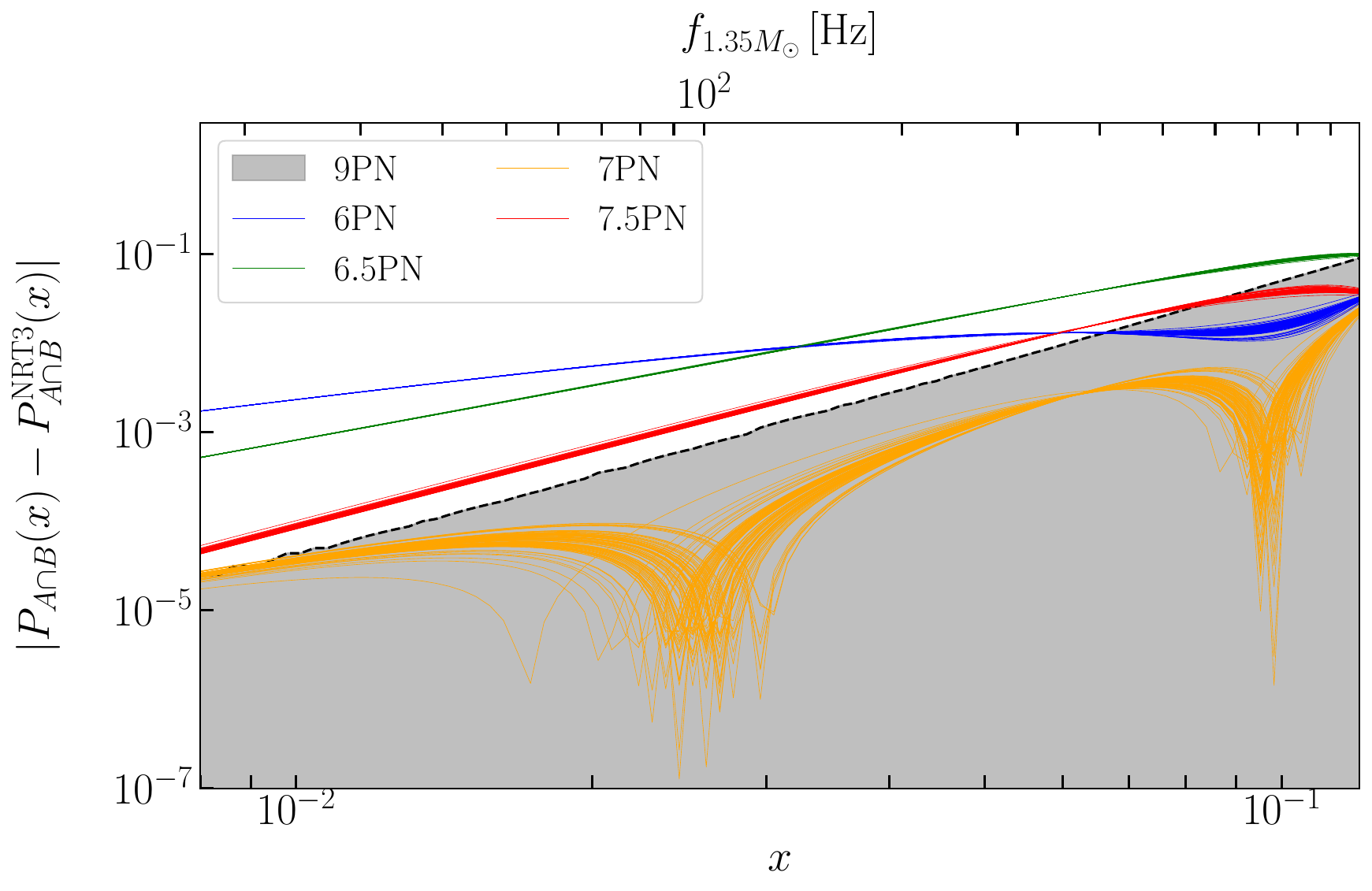}
\caption{The absolute error between the inferred $P_{A\cap B}^{\rm 9PN}$ and injected $P_{A\cap B}^{\rm NRT3}$ curves for $N=100$ events. The 90\% upper bound is marked by the dashed black line. The $P_{A\cap B}(x)$ for the 6PN, 6.5PN, 7PN and 7.5PN orders calculated using the catalog's source parameters are also plotted.}
\label{fig: combined_diffplot}
\end{figure}

\newcontent{We then take the absolute deviation of $P_{A\cap B}^{\rm 9PN}$ with respect to $P_{A\cap B}^{\rm NRT3}$ for each event, and employed the same method of combining the resulting posterior distributions of these deviations for all 100 events.} The upper 90\% bound of this error is shown in Fig.~\ref{fig: combined_diffplot}. We also plot the absolute deviation between $P_{A\cap B}(x)$ of known PN orders (i.e., up to 6PN, 6.5PN, 7PN, 7.5PN, using the source parameters in the catalog), and $P_{A\cap B}^{\rm NRT3}$. At lower frequencies, i.e., for $x<0.06$ ($f_{1.35M_\odot}\lesssim 300 \,\rm Hz$), the inferred $P_{A\cap B}^{\rm 9PN}$ achieves a smaller error than the 7.5PN order. \newcontent{It is also noticeable that for the late inspiral $x \sim 0.1$ ($f_{1.35M_{\odot}} \sim 800 \,\rm Hz$), the integer PN orders, in general, have smaller error than the half-integer PN orders.} This is due to the repulsive nature of the PN expression at half-integer order and the oscillatory nature of the PN expansion. Nevertheless, we see $P^{\rm 9PN}(x)$ to approach and even exceed the accuracy of the PN orders, especially at early inspiral and when we include more events in the calibration. We also note that we generally find a larger deviation in the results employing different PN orders compared to the spread of $P(x)$ (cf.~bands in Fig.~\ref{fig: combined_diffplot}). Hence, the presented data-driven approach can be employed to probe fundamental physical concepts directly, e.g., the PN expansion of tidal effects. Similar results, as the one presented in Fig.~\ref{fig: combined_diffplot}, can also be obtained for the 2L-ET configuration without noticeable differences, which agrees with the findings of~\cite{Branchesi:2023mws} that the exact ET configuration will have only a small effect on the obtained results. When combining ET with CE, we find smaller uncertainties but also become prone to biases in our parameter estimation, as one would expect due to the difference between the employed injection and recovery waveform model and the assumption of a common $P(x)$ for all events; cf.~\cite{supplementary_material}.

\begin{figure}[t]
\centering  \includegraphics[width=\linewidth]{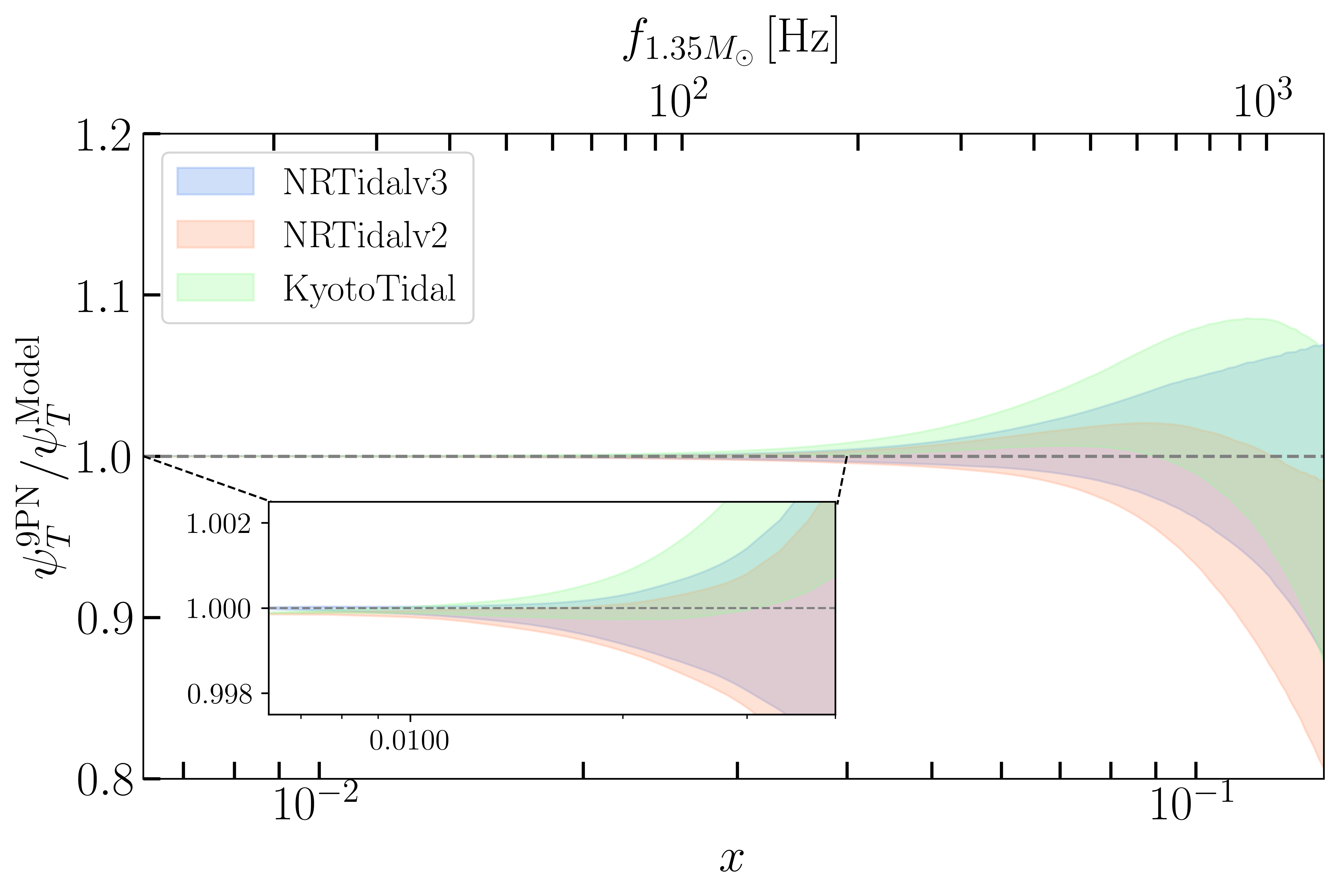}
\caption{The ratio $\psi_T^{\rm 9PN}/\psi_T^{\rm Model}$ between the tidal 9PN phase calculated using $P_{A,B}^{\rm 9PN}$ and that of other tidal models \nrtidalvthree, \nrtidalvtwo, and \kyototidal\ calculated using the catalog's source parameters.}
\label{fig: modelcomp}
\end{figure}

The entire tidal phase $\psi_T^{\rm 9PN}$ can also be computed using the inferred combined posteriors for $P_{A,B}^{\rm 9PN}(x)$, and compared against other existing tidal models, such as \nrtidalvthree~\cite{Abac:2023ujg}, \nrtidalvtwo~\cite{Dietrich:2019kaq}, and \kyototidal~\cite{Kawaguchi:2018gvj} calculated using the source parameters of the catalog. 
Taking the ratio $\psi_T^{\rm 9PN}/\psi_T^{\rm Model}$ between the waveforms, as shown in Fig.~\ref{fig: modelcomp}, \newcontent{we observe that the 90\% confidence interval for $\psi_T^{\rm 9PN}/\psi_T^{\rm Model}$ for \kyototidal\ already deviates from unity at low frequencies ($x \gtrsim 0.04$ or $f_{1.35M_{\odot}} \gtrsim 200\, \rm Hz$) while the one for \nrtidalvtwo\ marginally agrees at low frequencies but deviates at higher frequencies ($x \gtrsim 0.12$ or $f_{1.35M_{\odot}} \gtrsim 1000\, \rm Hz$). Meanwhile, the band agrees well with the injected \nrtidalvthree\ waveforms even at later frequencies}. These results highlight that already, with about one year of observational data, the presented data-driven approach becomes competitive with state-of-the-art techniques for modeling GW signals. Hence, using a conjunction of data-driven, analytical knowledge, and NR simulations yields great potential to further improve our understanding of the BNS coalescence and to accurately extract GW source properties. 

\paragraph{Outlook.---}
\label{section: Outlook}
We have presented a novel, data-driven method to construct phenomenological BNS waveform models by leveraging information about the tidal phase from binary neutron star coalescences using ET observations. \newcontent{The newly obtained waveform models could then be employed for a more accurate analysis of observational data to obtain tighter constraints on the NSs' tidal deformabilities, and consequently on the EOS governing supranuclear-dense matter.} We have demonstrated the feasibility of this approach by combining data from one hundred simulated ET detections to calibrate and constrain unknown higher-order coefficients in a pseudo-PN tidal waveform. This number of detections with the employed SNRs mimics roughly one year of operation for ET.

Our approach assumes that the BBH baseline waveform is accurate and well modeled, and that the tidal contributions are dominated by the (2,2)-mode. In the future, the method could be further improved by including higher-order modes in the tidal contributions, e.g., by rescaling the inspiral part proportional to the $m$-mode, and through identification and mitigation of systematic uncertainties in the waveform models~\cite{Gamba:2020wgg,Kunert:2021hgm}, which become more significant for high-SNR events, and, for instance, with the inclusion of more complicated physics. Overall, our method complements traditional approaches for waveform modeling and provides a new avenue for future developments. Used in synergy with existing methods, one will hopefully be able to reach the necessary accuracy to make full use of the next generation of GW detectors. Such an improvement will then enable us to provide accurate information on astrophysical, cosmological, and nuclear-physics properties, e.g., through well-measured neutron star masses, refined knowledge about the EOS of neutron stars, and improved constraints on the Hubble constant.  \\

\paragraph{Acknowledgments.--} This material is based upon work supported by the NSF's LIGO Laboratory which is a major facility fully funded by the National Science Foundation. We thank Hauke Koehn for providing the relevant EOS set for catalog construction, and we also thank Henrik Rose and Thibeau Wouters for providing helpful comments. The parameter estimation runs were performed using the hypatia cluster at the Max Planck Institute for Gravitational Physics.
We acknowledge funding from the Daimler and Benz Foundation for the project ``NUMANJI" and from the European Union (ERC, SMArt, 101076369). Views and opinions expressed are those of the authors only and do not necessarily reflect those of the European Union or the European Research Council. Neither the European Union nor the granting authority can be held responsible for them. 

\bibliography{References}

\onecolumngrid
\section{Supplementary Material to Data-driven approach for extracting tidal information from neutron star binary mergers observed with the Einstein Telescope}
\section{I. Employed Waveform Models}
\label{section: supmat I}

The phase in the \nrtidalvthree\ model is given by 
\begin{equation}
    \psi_T^{\rm NRT3} = -c_{\rm Newt}^A \kappa_A(\hat{\omega}) x^{5/2} P_A^{\rm NRT3}(x) + \left[A \leftrightarrow B \right],
\end{equation}
where $x = (\hat{\omega}/2)^{2/3} = (\pi Mf)^{2/3}$ is the PN parameter (or frequency), the coefficient $c_{\rm Newt}^A$ depends on $X_{A,B}$, $\kappa_A(\hat{\omega})$ is the dynamical tidal parameter, and $P_{A}^{\rm NRT3}(x)$ is a Padé approximant ansatz. Some of the coefficients of $P_{A}^{\rm NRT3}(x)$ are selected to be constrained by the 7.5PN, while the rest of the coefficients are calibrated to a large set of NR simulations, including high-mass-ratio systems across various EOS.

In this work, we assume a 9PN extension to the analytical 7.5PN tidal phase (which we attach to the \imrphenomxas\ BBH model, obtaining \imrphenomxasninePN)
\begin{equation}\label{eq:phase9pn}
\begin{split}
   \psi_T^{\rm 9PN} = &-c_{\rm Newt}^A \kappa_A(\hat{\omega}) x^{5/2} P_A^{\rm 9PN}(x; c_3^{A}, c_{7/2}^{A}, c_4^{A})+ \left[A \leftrightarrow B\right], 
\end{split}
\end{equation}
with the rational function $P_A^{\rm 9PN}$ being a polynomial given by
\begin{equation}\label{eq: 9PNtidal}
\begin{split}
    P_A^{\rm 9PN}(x; c_3^{A}, c_{7/2}^{A}, c_4^{A}) = 1 + c_1^A x + c_{3/2}^Ax^{3/2} + c_2^A x^2 + c_{5/2}^A x^{5/2} + c_3^A x^3 + c_{7/2}^Ax^{7/2} + c_4^A x^4.
\end{split}
\end{equation}
The six additional free parameters ($c_3^{A,B}, c_{7/2}^{A,B}, c_4^{A,B}$) are the HOCs, which we infer via parameter estimation (PE). Note that these coefficients do not necessarily have to be identical to the ones that would be derived purely from the PN formalism: this extension is treated as a \textit{pseudo}-PN tidal phase.

\begin{figure}[h]
\centering  \includegraphics[width=0.328\linewidth]{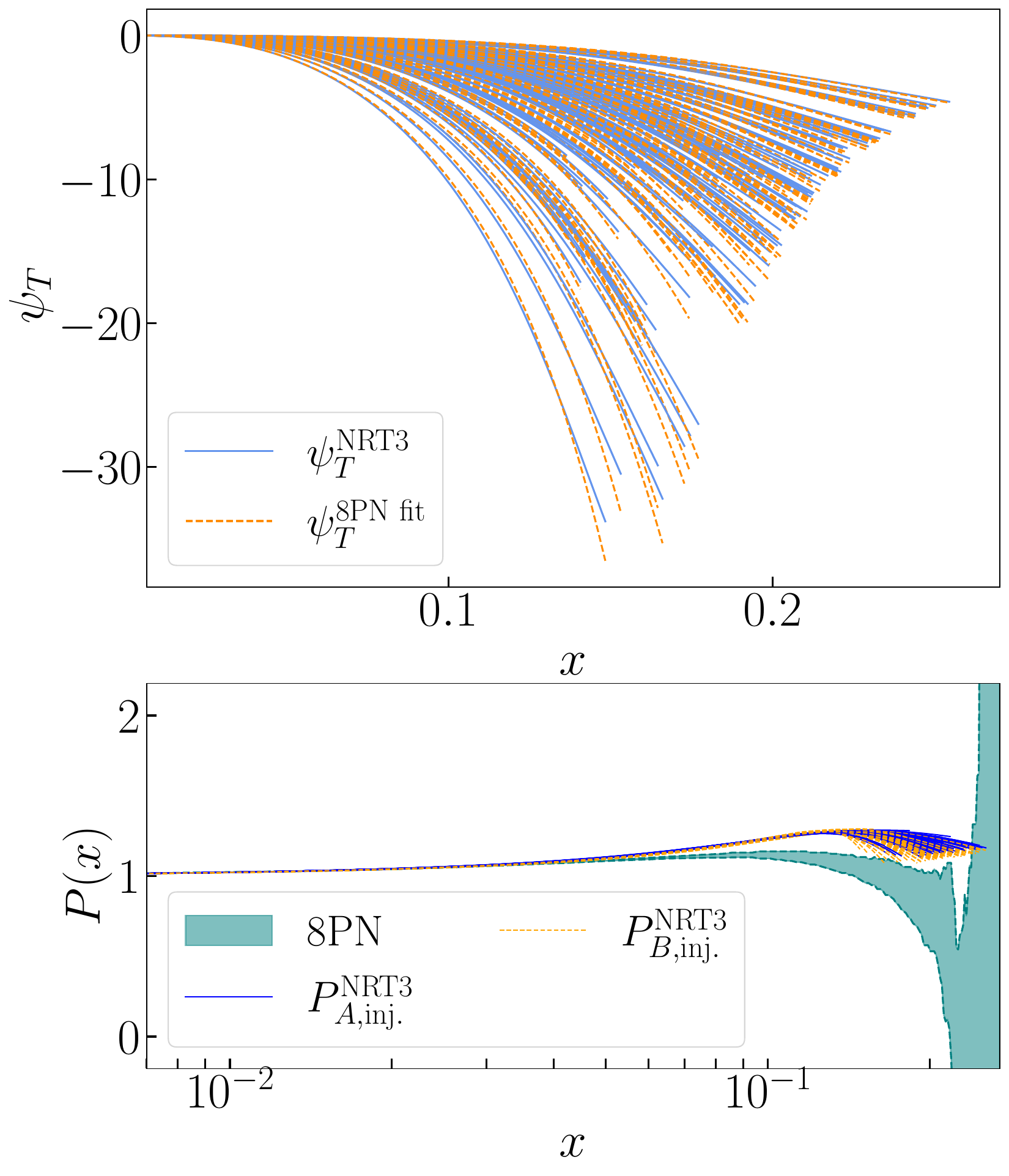}
\centering  \includegraphics[width=0.328\linewidth]{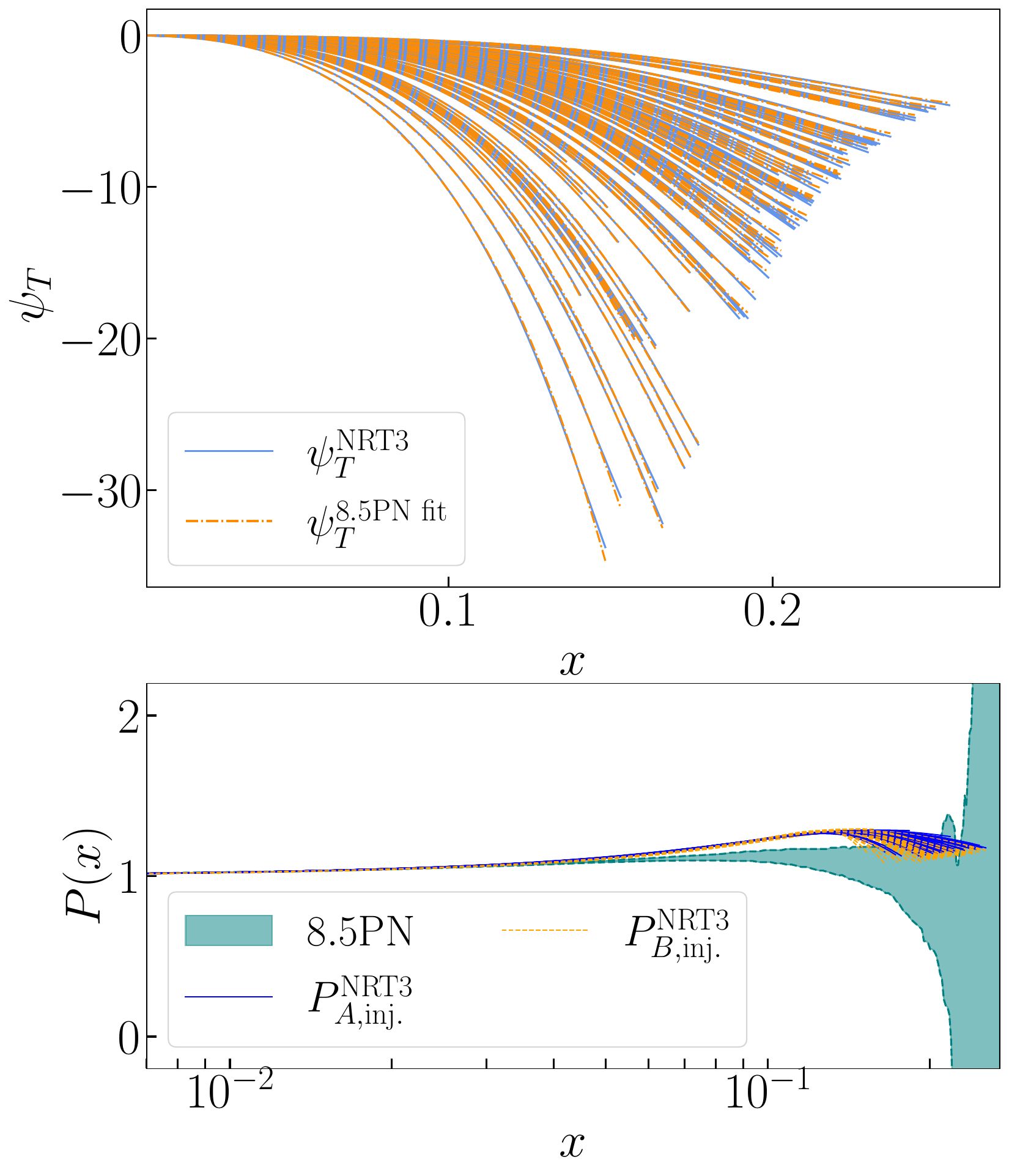}
\centering  \includegraphics[width=0.328\linewidth]{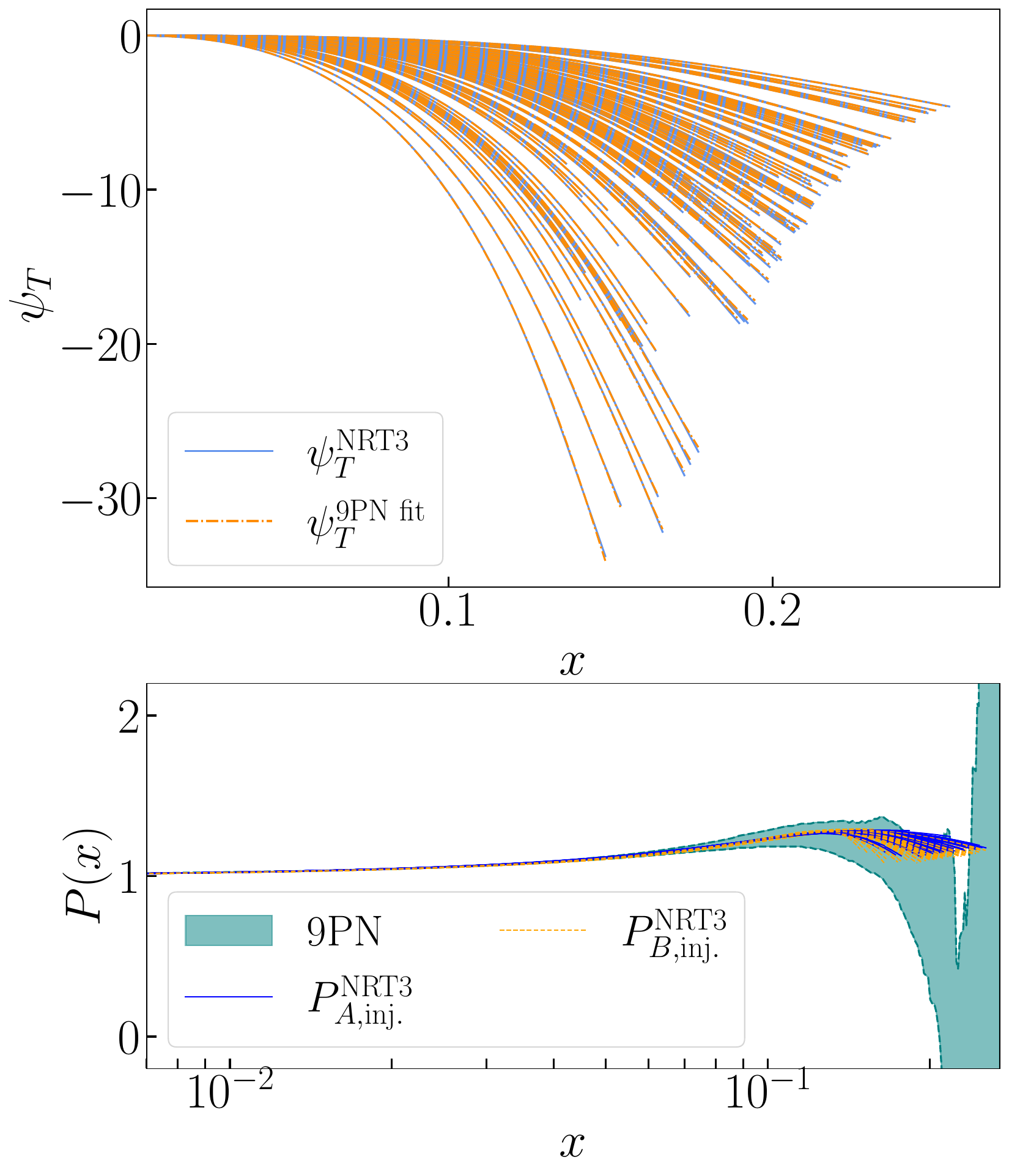}
\caption{\newcontent{{\textit{Upper panels}: Linear regression fits of the PN phase at 8PN, 8.5PN, and 9PN orders for all injections with respect to $\psi_T^{\rm NRT3}$. \textit{Lower panels}: The corresponding 90\% confidence bands of the joint posteriors of the different PN orders using the $\Delta$-ET configuration, together with the injected \nrtidalvthree\ curves. We observe the 9PN tidal extension to sufficiently capture the injected information from \nrtidalvthree.}}}\label{fig: PNorderfits}
\end{figure}

\begin{figure}[h]
\centering\includegraphics[width = 0.5\linewidth]{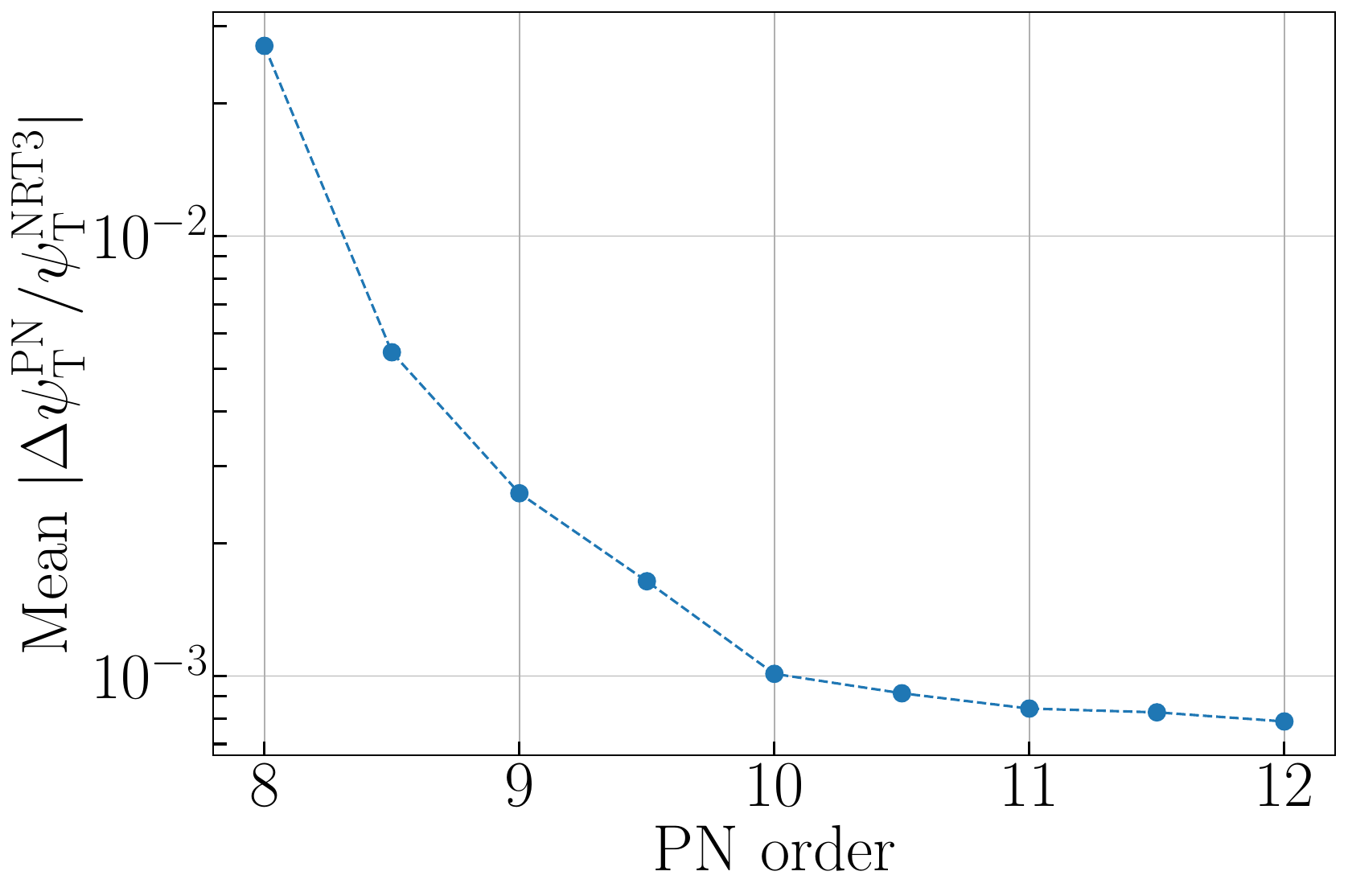}
\caption{\newcontent{The mean absolute relative error between the PN phase with respect to $\psi_T^{\rm NRT3}$ at different PN orders. We observe a decrease in the mean error with increasing PN order.}}\label{fig: PNmeanerror_comp}
\end{figure}

\newcontent{We find that the 9PN-extension to the tidal phase is both sufficient and computationally efficient enough to demonstrate our approach. To verify this, we employ the same approach of combining joint $P(x)$ posteriors using pseudo-8PN ($\mathcal{O}(x^{3})$) and 8.5PN ($\mathcal{O}(x^{7/2})$) extensions to the tidal description, and we show in Fig.~\ref{fig: PNorderfits} that the 8PN and 8.5PN extensions alone are insufficient in capturing the injected tidal information from \nrtidalvthree, even at $x\lesssim 0.1$, though their confidence bands are narrower than in the 9PN case. This is also evident from the results of linear regression fitting of the PN coefficients (upper panels of Fig.~\ref{fig: PNorderfits}) to the \nrtidalvthree{} phase, where the 9PN phase fits better compared to the 8PN and 8.5PN phases. This is also supported by the overall smaller mean error for 9PN than the 8PN and 8.5PN orders, shown in Fig.~\ref{fig: PNmeanerror_comp}.}

\newcontent{We also extend the linear regression fits to \nrtidalvthree\ beyond 9PN and up to 12PN order, and calculate the absolute relative error $|\Delta\psi_T^{\rm PN}/\psi_T^{\rm NRT3}| = |(\psi_T^{\rm PN}-\psi_T^{\rm NRT3})/\psi_T^{\rm NRT3}|$ between the fits of each PN order and the corresponding \nrtidalvthree\ phase (up to the lowest common merger frequency of all events). The mean $|\Delta\psi_T^{\rm PN}/\psi_T^{\rm NRT3}|$ as a function of PN order are shown in Fig.~\ref{fig: PNmeanerror_comp}, where we observe a decrease of the error as the PN order increases. At 9PN order, we already reach a mean error of less than 0.3\%. With this accuracy achieved by the 9PN order, and given that increasing the PN order also increases the inefficiency of the sampling in the PE runs due to addition of two unknown coefficients per PN order, we chose the 9PN extension to be the balance between accuracy and computational efficiency.}

\section{II. Prior Choices and Likelihood Settings}
\label{section: supmat II}
\begin{table}[htb]
\centering
\caption{\label{table: prior} Injection and recovery prior distributions for the different source parameters for \imrphenomxasninePN. The distributions used for the injection are the same as in the priors unless otherwise stated. Most parameters are represented by uniform distributions $\mathcal{U}$. Note that $\mathcal{U}^{M_A}$ and $\mathcal{U}^{M_B}$ indicates uniform distributions for $M_A$ and $M_B$, respectively. $\mathcal{U}^{\rm AS}$ denotes uniform distribution in aligned spins, and $\mathcal{U}^{\rm CoVol}$ means uniform in comoving volume. }
\begin{tabular}{p{6cm}| p{5cm}} 
\hline \hline 
Parameter & Distribution\\
\hline

$M_{A,B}$ [$M_{\odot}$]  (injection only)       & $\mathcal{U}^{M_A} (1.0, 2.0)\mathcal{U}^{M_B} (1.0, 2.0)q^2$\\
$\mathcal{M}_c$ [$M_{\odot}$] (prior only)        & $\mathcal{U} (\mathcal{M}_c^{n; \rm inj} - 0.05, \mathcal{M}_c^{n; \rm inj} + 0.05)$ \\
$ 1/q $   (prior only)                 & $\mathcal{U}(0.5, 1.0)$                                                             \\
$\chi_{A,B}$                  & $\mathcal{U}^{\rm AS}(0, 0.15)$                         \\
 $D_L$ [Mpc] & $\mathcal{U}^{\rm CoVol}(10, 1000)$                                                 \\
  $\delta$             & $\rm Cosine$                                                                           \\
     $\alpha$      & $\mathcal{U}(0, 2\pi)$ (periodic)                                                   \\
 $\theta_{jn}$ & $\mathcal{U}(0, \pi)$                                                             \\
 $\psi_p$                    & $\mathcal{U}(0, \pi)$                                                               \\
 $\phi_c$                    & $\mathcal{U}(0, 2\pi)$ (periodic)                                                   \\
$\Lambda_{A,B}$ (prior only)               & $\mathcal{U}(0, 5000)$                                                               \\
$c_3^{A,B}$ (prior only)                  & $\mathcal{U}(-1000, 1000)$                                                        \\
$c_{7/2}^{A,B}$  (prior only)             & $\mathcal{U}(-4000, 8000)$                                                        \\
$c_4^{A,B}$  (prior only)                 & $\mathcal{U}(-15000, 4000)$   \\                 \hline \hline                
\end{tabular}
\end{table}

We show in Table~\ref{table: prior} the different distributions used for the injection of \imrphenomxasnrtidalthree\ and recovery of the source parameters of \imrphenomxasninePN.

\begin{figure}[h]
\centering\includegraphics[width = 0.6\linewidth]{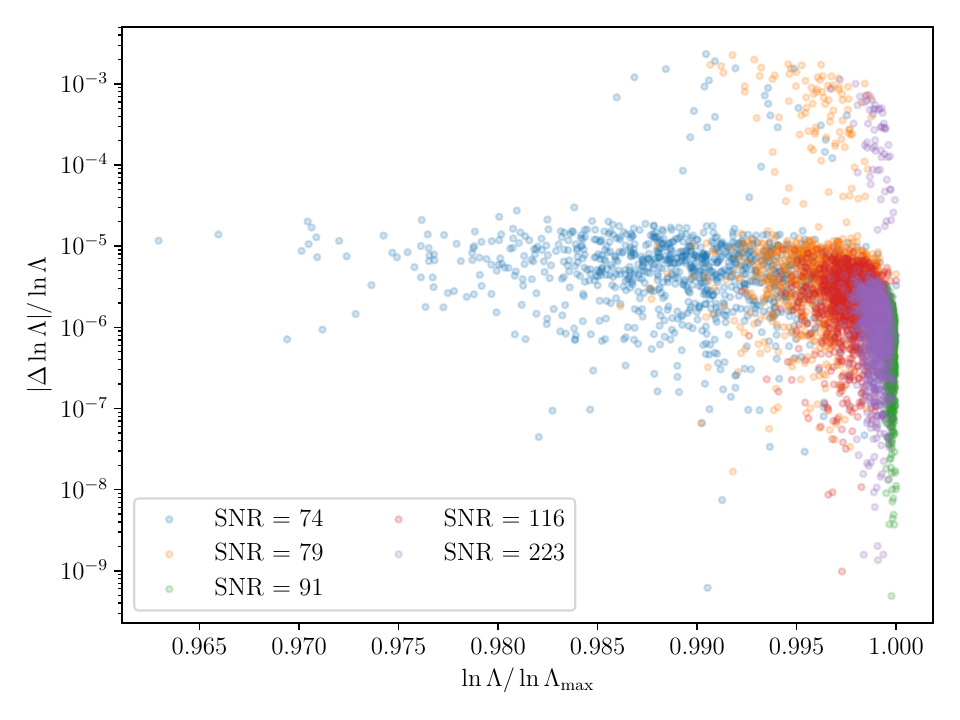}
\caption{\newcontent{The relative error of the multibanding likelihood as a function of the normalized non-multibanded likelihood. We see the relative errors to be consistently below $\mathcal{O}(10^{-2})$ for the events with different SNRs, and to be of the order of $10^{-5}$ or below for most cases.}}\label{fig: likelihood_comp}
\end{figure}

\newcontent{In this work, we employ an adaptive frequency resolution (also known as multibanding) method in the evaluation of the likelihoods for the \imrphenomxasninePN\ waveform, which increases the computational speed of our PE runs. To test the robustness of the method, we compare the log-likelihoods of five different events of various SNRs by taking the relative error of the multibanding likelihoods with respect to one without employing the multibanding technique, that is, $|\Delta \ln \Lambda|/\ln \Lambda = (\ln \Lambda_{\rm MB} - \ln \Lambda)/\ln \Lambda$. We show in Fig.~\ref{fig: likelihood_comp} this relative error as a function of the non-multibanded likelihood normalized to its maximum value, and demonstrate the accuracy of this method to $\mathcal{O}(10^{-5})$ in most of the points, even for high-SNR events. }

\section{III. 2L-ET and ET\&CE network}
\label{section: supmat III}
\begin{figure}[h]
    \centering  \includegraphics[width=0.328\linewidth]{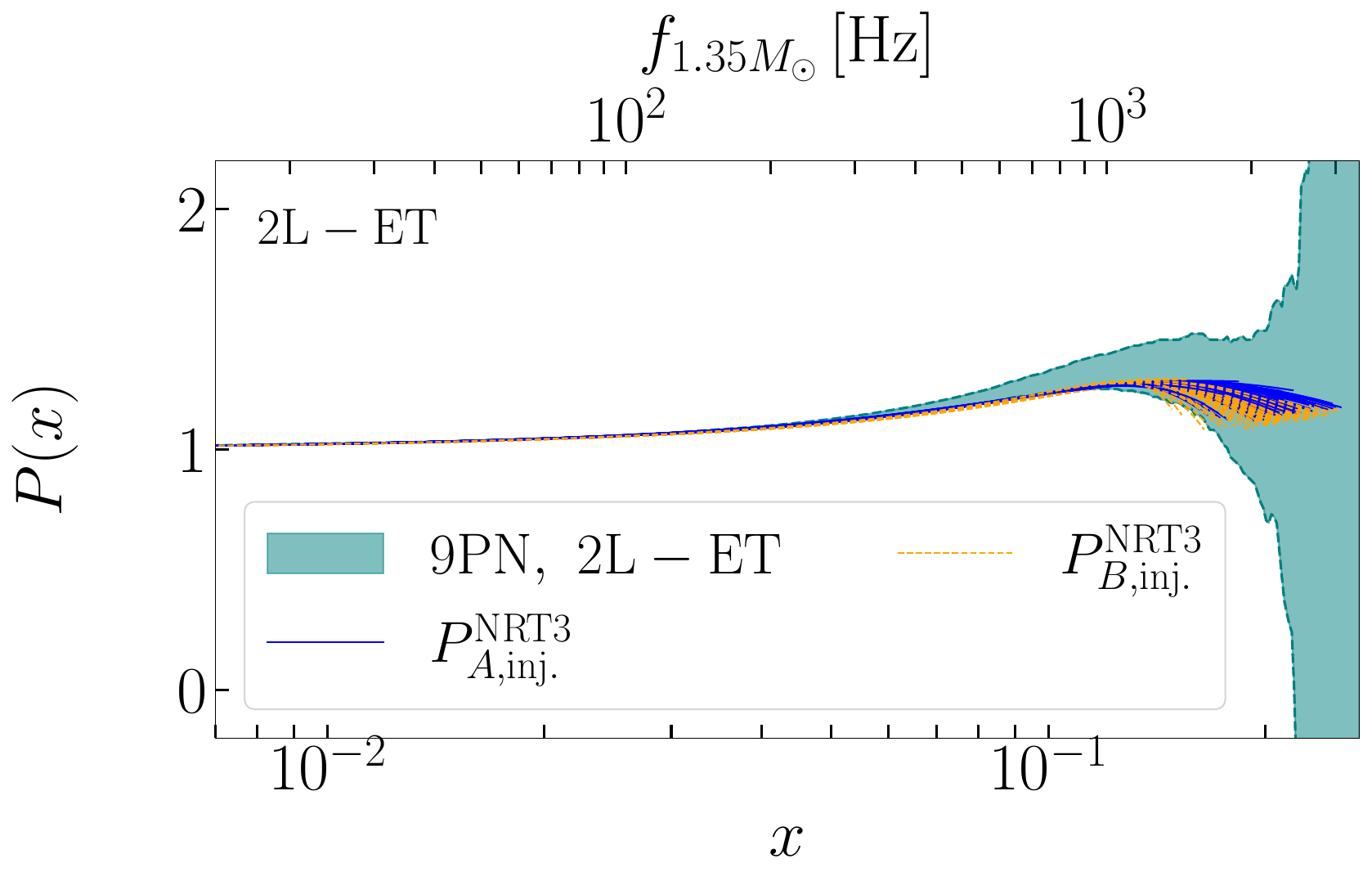}
    \centering  \includegraphics[width=0.328\linewidth]{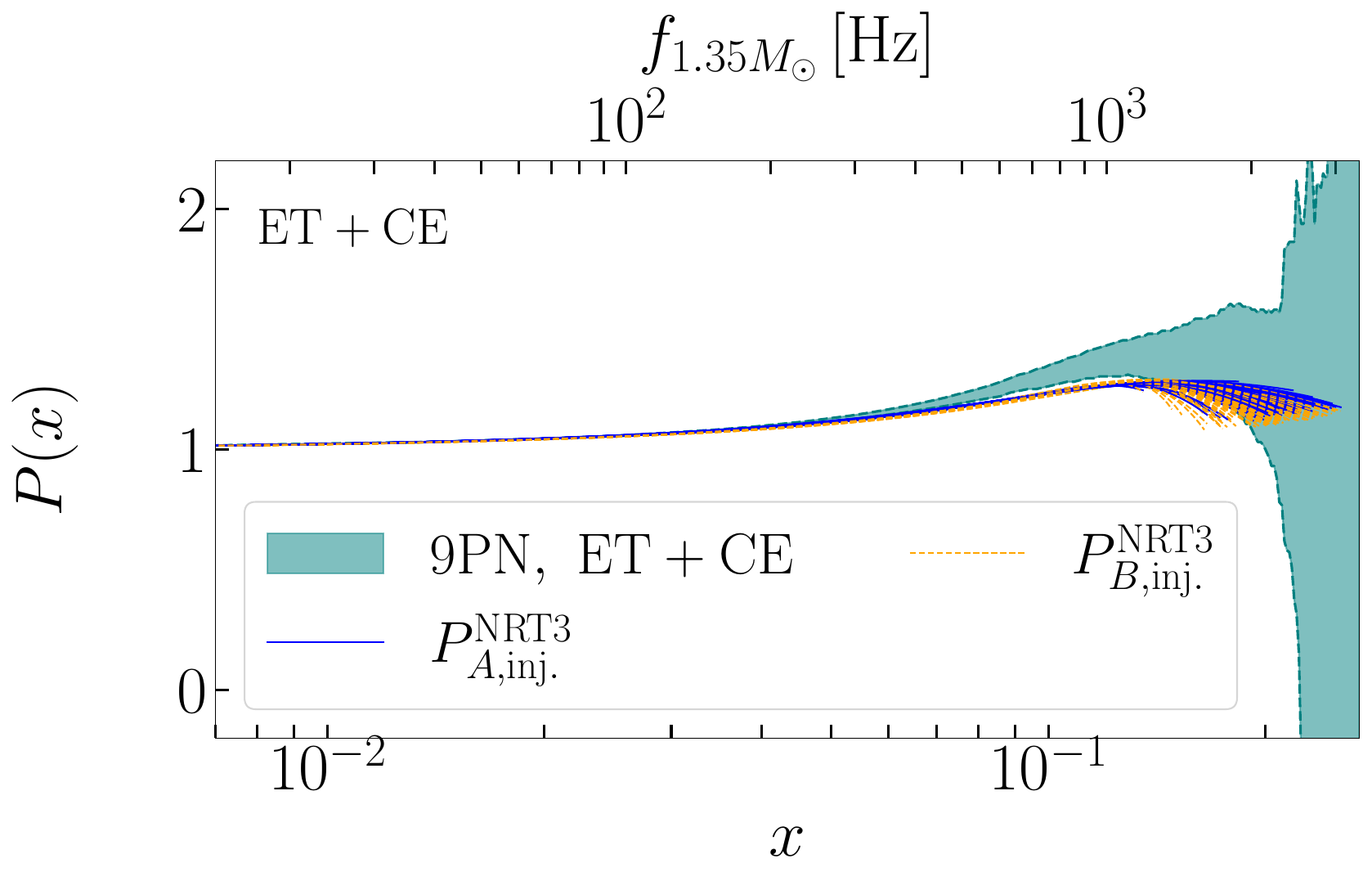}
    \centering  \includegraphics[width=0.328\linewidth]{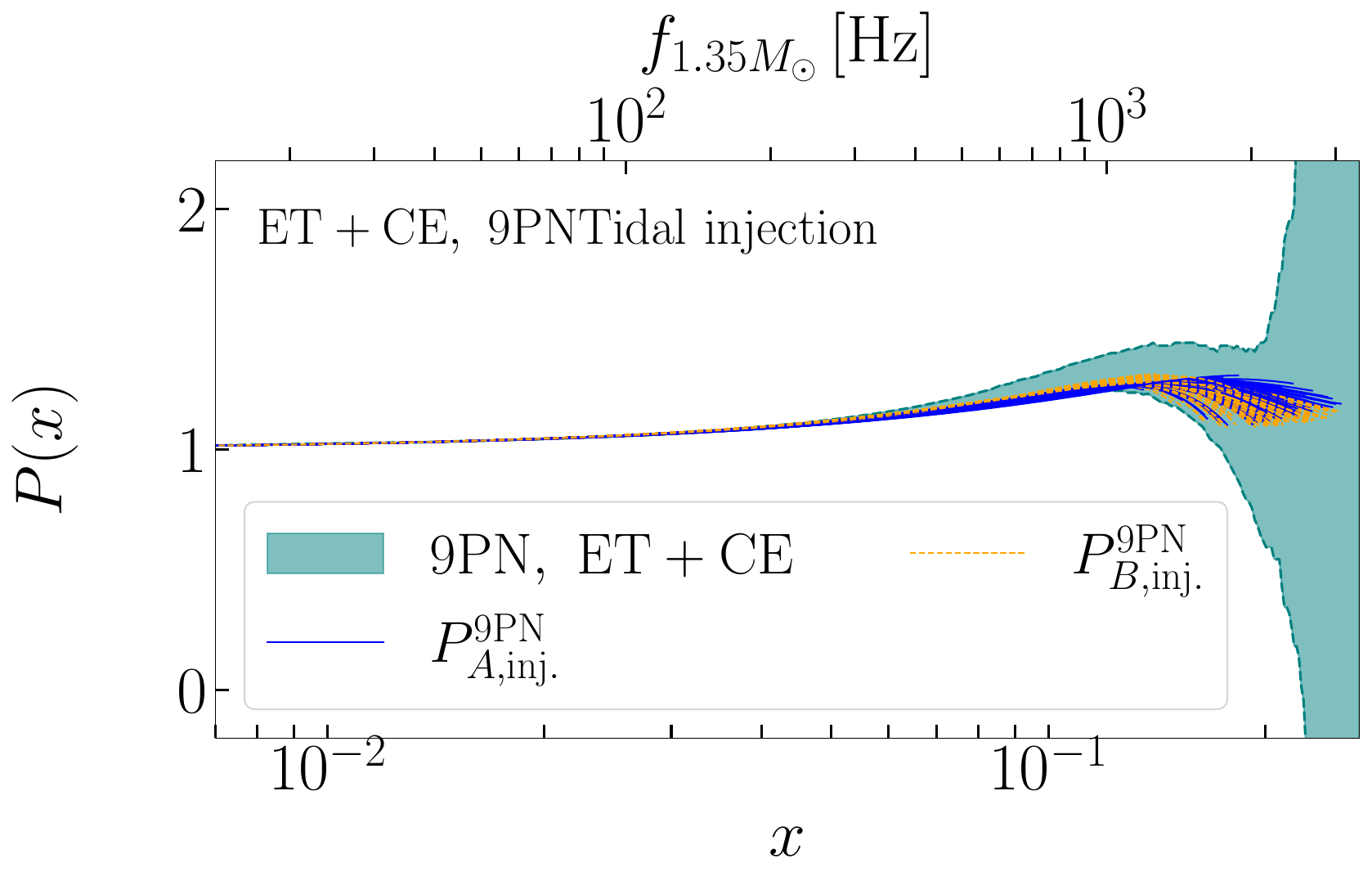}   
    \caption{\textit{Left panel}: 
    The confidence band of the joint posterior of $P^{\rm 9PN}(x)$ for $N = 100$ events using the 2L-ET configuration, together with the injected \nrtidalvthree\ curves. We observe a slight shift of the band towards larger values of $P(x)$. \newline \textit{Middle panel}: The confidence band of the joint posterior of $P^{\rm 9PN}(x)$ for $N = 100$ events using the ET+CE configuration, together with the injected \nrtidalvthree\ curves. We also observe a (larger) shift of the band towards larger values of $P(x)$. \newline \textit{Right panel}: The confidence band of the joint posterior of $P^{\rm 9PN}(x)$ for $N = 100$ events using the ET+CE configuration. In this case, we inject and recover with the same \imrphenomxasninePN\ model, but this time, the dynamical tidal effects and spin dependence are excluded, as well as the mass-ratio dependence of the PN coefficients. No bias is visible, indicating that the source of the bias is due to the difference between the injection and recovery waveform models.}\label{fig: ET+CE100events}
\end{figure}

In this section, we add the results for our combined posteriors using an ET in the 2L configuration. We observe in Fig.~\ref{fig: ET+CE100events} (left panel) similar results to the $\Delta$ configuration, but with the inferred confidence band for $P(x)$ slightly shifted towards larger values with respect to the injected curves.  

For an ET+CE configuration, we find an even larger shift with respect to the injected values, as shown in the middle panel of Fig.~\ref{fig: ET+CE100events}. We attribute this bias primarily to systematics in the employed waveform models. More explicitly, we employ \imrphenomxasnrtidalthree\ for the creation of the mock data, but the pseudo-PN tidal model \imrphenomxasninePN\ for recovering the injected data. The finite accuracy of the 9PN representation to model the NRTidalv3 tidal effects will necessarily lead to small biases that will increase when events get combined. Using a higher PN expansion, with additional parameters, would reduce this effect but, on the other hand, increase the dimensionality of the problem and to less stringent constraints. For this reason, we have decided to stick to a 9PN representation, while further quantifying the necessary pseudo-PN order that one has to employ for a given number of detections would be needed. 

In addition, other effects (presumably less dominant) could also introduce small biases, e.g., the selection bias associated with including only the events with the highest SNRs, which the ET+CE detectors are more likely to detect. Furthermore, this can also be due to correlations between the source parameters (i.e., the masses) and the HOCs. We note that the known PN coefficients up to 7.5PN already have an explicit $X_{A,B}$-dependence, and the exact relationship between the true 9PN coefficients and $X_{A,B}$ remains unknown.

To validate our results, we perform injection-recovery runs for all events with $\imrphenomxasninePN$. However, we restrict the model by excluding dynamical tidal effects and spins, as well as restricting the mass ratio dependence of the known PN coefficients to $q = 1.0$. 
This way, we avoid biases due to the different waveform models, reduce the possible influence of a non-constant $P(x)$ for different sources of the simulated population, and minimize possible selection biases influencing dynamical tides and spins-tidal effects.
Using the same methodology used previously, we combine the recovered posteriors for $P_{A,B}^{9\rm PN}$ to a posterior that describes the injected values well (see right panel of Fig.~\ref{fig: ET+CE100events}).  Accounting for all the sources of bias is a non-trivial task, which would require a large number of additional PE runs, which we reserve for a future study. 

\end{document}